\documentstyle[12pt,espcrc1,epsfig]{article}
\large
\newcommand{\be}{\begin{eqnarray}}
\newcommand{\ee}{\end{eqnarray}}
\newcommand\del{\partial}

\begin{document}
\setlength{\baselineskip}{21pt}
\pagestyle{empty}  
\vfill                                                                          
\eject                                                                          
\begin{flushright}                                                              
SUNY-NTG-97/8

\end{flushright}                                                                
                                                                                
\vskip 1.0cm 
\centerline{\Large
Fermion determinants in matrix models of QCD }
\vskip 0.5cm 
\centerline{\Large  at nonzero chemical potential}

\vskip 1.0 cm 
                                                                 
\centerline{M.A. Halasz$^{\rm a}$,  A.D. Jackson$^{\rm b}$ 
and J.J.M. Verbaarschot$^{\rm a}$}
\vskip 0.2cm            
\centerline{$^{\rm a}$ Department of Physics, SUNY, Stony Brook, 
New York 11794, USA}
\centerline{$^{\rm b}$ Niels Bohr Institute, Blegdamsvej 17, Copenhagen,
DK-2100, Denmark}
\vskip 1.3cm

\centerline{\bf Abstract}
The presence of a chemical potential completely changes the analytical
structure of the QCD partition function. In particular, the eigenvalues
of the Dirac operator are distributed over a finite area in the complex 
plane, whereas the zeros of the partition function in the complex mass 
plane remain on a curve.
In this paper we study the effects of the fermion determinant at nonzero 
chemical potential  on the
Dirac spectrum by means of the resolvent, $G(z)$,
of the QCD Dirac operator. The 
resolvent is studied both in a one-dimensional $U(1)$ model (Gibbs model) 
and in a random
matrix model with the global symmetries of the QCD partition function.
In both cases we find that, if the argument $z$ of the resolvent is not equal to
the mass $m$ in  the fermion determinant, 
the resolvent diverges in the thermodynamic
limit. However, for $z =m$ the resolvent in both models is well defined. 
In particular, the
nature of the limit $z \rightarrow m$ is illuminated in the Gibbs model.  
The phase structure of the random matrix model in the complex $m$ and 
$\mu$-planes is investigated  both  by a saddle point approximation and
via the distribution of Yang-Lee zeros. Both methods are in complete agreement
and lead to a well-defined chiral condensate and quark number density.  
\vfill                                                                          
\noindent                                                                       
\begin{flushleft}
March 11, 1997
\end{flushleft}
\eject
\pagestyle{plain}
\noindent
\section{Introduction}
\vskip 0.5cm
During the past decade, numerical lattice QCD simulations have provided
a great deal of insight regarding strong interactions
at finite temperature (see \cite{DeTar,Ukawa} for a review). 
In spite of steady progress \cite{barbour}
numerical simulations of QCD at finite baryon density have proved to be
remarkably difficult and their results remain inconclusive
\cite{everybody,Kogut}. 
The main reason for these difficulties can be traced back to the way
the quark number chemical potential enters
a Euclidean field theory. It results in  
a fermion determinant and a probability measure 
that are not 
positive definite, making standard Monte Carlo simulations impossible.
A possible solution to this problem is the quenched approximation, i.e.,
ignoring the fermion determinant in generating gauge field configurations.
This method has been quite successful in other applications of LGT (see e.g., 
\cite{Weingarten}). 
Unfortunately, results obtained 
for finite chemical potential show a critical value of $\mu$ 
at half the pion mass instead of a third of the nucleon mass \cite{everybody}. 
Two possible explanations have
been offered in the literature: i) The problems are an artifact of 
relatively small lattices and would be cured in the continuum
limit \cite{Bilic,Vink}. 
ii) The quenched approximation is responsible for these unphysical results
 \cite{Gibbs1,Gibbs}.
Regarding the latter point we mention one particularly 
interesting suggestion \cite{Gocksch,Stephanov2}, 
namely that the quenched approximation is obtained as the limit of the 
number of flavors going to zero with an equal number of fermions and 
conjugate fermions, i.e., the limit of a partition function in which
only the absolute value of the determinant enters. Recently, this suggestion
has been made much more explicit by Stephanov \cite{Stephanov2}
within the framework of
a random  matrix model with the chiral and flavor structure of the 
Dirac operator (see \cite{review} for a review).  
He has shown {\em analytically\/} that the quenched Dirac 
spectrum is obtained in the limit as both the number of quarks and
conjugate quarks tend to zero. 
Remarkably, the chiral condensate of this model (with the full determinant 
included)
is independent of the number of flavors if $N_f$ is a positive 
integer. However, since $real$ QCD does not have such conjugate quarks, it 
seems unlikely that its $N_f \rightarrow 0$ limit would lead to the 
quenched approximation.

This leads to the more general question regarding the extent to which 
the quenched approximation can be trusted. Another familiar example of its 
failure is the chiral
limit, in which the quenched theory does not reproduce the correct chiral
logarithms \cite{Golterman}. The failure of the quenched approximation
is most obvious if the quark mass is less than the smallest nonzero 
Dirac eigenvalue. 
Then the chiral condensate 
in the full theory is zero for two or more flavors, finite for one flavor and
possibly divergent in the quenched approximation \cite{teper}. 
The reason for the failure is
that the leading contribution to the expectation value of an operator is
a subleading contribution to the partition function. 
A similar mechanism is at work
in one-dimensional $U(N)$ lattice models \cite{Gibbs,Damgaard}. There, the
leading contribution to the fermion determinant cancels after averaging over 
the gauge group. This is no longer true if the gauge group is $SU(N)$. Then,
 the chiral condensate in the quenched and unquenched theory are indeed equal
\cite{damlett,Bilic}.   

The random matrix model studied in this paper
is the extension to nonzero chemical potential of the model 
introduced in \cite{chrmt}.
The partition function is defined by
\be
Z= \int DH P(H){\det}^{N_f}(m-D) \ , 
\label{rmt}
\ee 
where $D$ is the random Dirac operator
\be
D= \left ( \begin{array}{cc} 0 & iW+\mu \\ iW^\dagger +\mu & 0 
\end{array} \right ) \ ,
\label{Dirac}
\ee
and $W$ is a complex matrix distributed according to the 
Gaussian probability distribution
distribution $P(W)$
\be
P(H) = \exp(-n\Sigma^2 {\rm Tr} WW^\dagger) \ .
\label{prob}
\ee
For nonzero chemical potential, the Euclidean QCD Dirac operator is
$
D= \gamma_\mu (\del_\mu + i A_\mu) + \gamma_0 \,\mu
$.
The random matrix partition function is 
obtained from the QCD partition function by 
replacing the $\mu$-independent terms of  the 
matrix elements of the Dirac operator in a chiral basis
by Gaussian random variables. In this basis, the term $\mu \gamma_0$ results 
in the term $\mu$ times the identity in the off-diagonal blocks in 
(\ref{Dirac}). We wish to emphasize that this random matrix model is
a $schematic$ model of the QCD partition function.

The random matrix model (\ref{rmt}) is in the class of nonhermitean random 
matrix models. Recently, a variety of 
such models with applications to different physical problems have been discussed
in the literature (see for example \cite{nonhermitean,zee}).

The Euclidean Dirac operator at non-zero chemical potential is non-Hermitean, 
and, in general, its eigenvalues will be distributed in the 
complex plane.
Indeed, this was found in lattice simulations by the Urbana group 
\cite{everybody}.
The quenched case is defined as the theory which is obtained by ignoring the
fermion determinant in the partition function (\ref{rmt}). 
In \cite{Stephanov2} it was 
pointed out that this theory is not obtained by taking the limit 
$N_f\rightarrow 0$ in (\ref{rmt}), but rather from the partition function  
(\ref{rmt}) with ${\det}(m-D)^{N_f}$ replaced
by $|{\det}(m-D)|^{N_f}$.

The partition function (\ref{rmt}) will be studied with the help of the
resolvent defined by 
\be
G(z) = \frac 1N\langle {\rm Tr}\frac 1{z-D}\rangle
\label{resolvent}
\ee
where the average is over the probability distribution (\ref{prob}) and
the fermion determinant.
For $z=m$, the singularities of the resolvent will be cancelled by the
fermion determinant in (\ref{rmt}). This is the physical (unquenched) case.

In this paper we will investigate the resolvent for $z\ne m$ 
at nonzero chemical potential. 
Our main point is that quenching completely changes the 
result for $G(z)$. In particular, we hope to convince the reader that, 
below a critical value of $z$, the thermodynamic limit of $G(z)$ does not 
exist if the valence quark mass is (i.e., $z$) is different from the sea
quark mass $m$. We shall show this for the $U(1)$ model introduced by Gibbs
\cite{Gibbs}. For the random matrix model (\ref{rmt})
the resolvent can be obtained analytically only for $z= m$. In this
case, it  is well defined and displays a first-order phase transition
at a nonzero critical value  of $\mu$. 
However, if valence and sea quark masses are different,   
numerical evidence shows that the thermodynamic limit of the
resolvent is divergent in this model as well. 

We start with some general definitions and discuss the relations
between the Dirac spectrum and the zeros of the partition function.  
Two simple models that lead to nonanalyticities in
$G(z)$ are discussed in section 3.
In section 4 we analyze the resolvent in the Gibbs model \cite{Gibbs1,Gibbs}.
In section 5 the random matrix model (\ref{rmt}) is analyzed by means of a
saddle point approximation and via the Yang-Lee zeros of the partition 
function (some of these results have been published in \cite{frank}). 
The quark number density is discussed in section 6. Numerical
results for the random matrix resolvent with $z\ne m$ are presented in section
7, and concluding remarks are made in section 8. 

\vskip 1.5cm
\noindent\section{Generalities} 
\vskip 0.5cm
We will study the spectrum of the Dirac operator via the analytic properties of
the resolvent (\ref{resolvent}).
All eigenvalues of the Dirac operator (\ref{Dirac}) occur in pairs $\pm 
\Lambda$. Therefore, the resolvent satisfies
\be
G(z) = -G(-z) \ .
\label{sym1}
\ee
If the distribution of the phases of the matrix elements of the Dirac operator
over the ensemble is reflection symmetric with respect to the real
axis, we also have the relation
\be
G(z^*) = G^*(z) \ .
\label{sym2}
\ee
Contrary to (\ref{sym1}) this relation is in general 
only valid after averaging.
For a Hermitean Dirac operator, both
relations are valid before averaging. 

Our central object of interest is the chiral 
condensate $\Xi$ defined by
\be
\Xi = \lim_{m\rightarrow 0} \Xi(m) \ ,
\label{condens}
\ee
where
\be
\Xi(m) = \lim_{N\rightarrow \infty}
\frac 1{NN_f} \del_m \log Z(m) \ .
\ee
The differentiation with respect to $m$ can be carried out before or after 
averaging over the gauge field configurations. In the first case, we 
factorize the fermion determinant as
\be
\det (D-m) = \prod_k(m-\lambda_k)
\ee
resulting in
\be
\Xi(m) = \frac 1N \sum_k \frac 1{m-\lambda_k} \equiv G(m) \ .
\ee
According to (\ref{sym1}), a nonzero chiral condensate necessarily implies that
$G(m)$ shows a discontinuity at $m=0$ along the real axis. 

In the second case, we factorize the partition function as
\be
Z(m,\mu) = \prod_k (m-m_k) \ ,
\ee
where the $m_k$ are the Yang-Lee zeros of the partition function.
In this case
\be
\Xi(m) = \frac 1{N_f N} \sum_k \frac 1{m-m_k} \ .
\ee
In the phase of broken chiral symmetry, $\Xi(m)$ shows a discontinuity
at $m =0$. This necessarily implies that the zeros form a cut in 
the thermodynamic limit.

The quark number density is defined as
\be
n_q = \frac 1{N\, N_f} \del_\mu \log Z \ .
\label{quarkdens}
\ee 
Here, too, we can differentiate before or after averaging over the
gauge field configurations. In the first case, the quark number density 
is given by
\be
n_q = \frac 1{N\,N_f} \sum_k \langle \frac 1{\mu -\zeta_k}\rangle \ ,
\ee
where the $\zeta_k$ are the eigenvalues of $\gamma_0(D+m)$.
They are the analogue of the eigenvalues of the propagator matrix 
introduced by Gibbs \cite{Gibbs1}.
If, on the other hand, we first average over the gauge field configurations,
the partition function can be factorized as
\be
Z(m,\mu) = \prod_k(\mu-\mu_k)
\ee
and results in the quark number density
\be
n_q = \frac 1{N\,N_f} \sum_k \frac 1{\mu -\mu_k} \ . 
\ee

Physically, we expect that the quark number density is zero below a certain 
nonzero critical value of $\mu$ determined by the baryon mass. This can be
achieved if the eigenvalues $\zeta_k$ are distributed with axial symmetry
in an annulus in the complex plane. Indeed, this is what has been observed 
numerically for QCD by Gibbs. 
Generally, we expect \cite{Shrock}  that the zeros of
the partition function are located on a one-dimensional manifold in the
complex plane. In order to have a zero baryon density below $\mu_c$, a 
natural expectation is that the zeros are distributed homogeneously 
along a circle with radius $\mu_c$.

The resolvent is analytic in the upper complex half plane for a Hermitean
Dirac operator. For nonvanishing chemical potential, the eigenvalues are 
scattered in the complex plane. 
Denoting the real and imaginary parts of the eigenvalues by
$\lambda_r$ and $\lambda_i$, the eigenvalue density is characterized by a
two dimensional spectral density
$ \rho(\lambda_r, \lambda_i)$. It is normalized according to 
\be
\int d\lambda_r d\lambda_i \rho(\lambda_r, \lambda_i) = 1 \ .
\ee
Using the fact that 
\be
\del_{\bar z} \frac 1z = \pi \delta^2(z) \ ,
\ee
where the complex delta function is defined as
$\delta^2(z) = \delta({\rm Re}(z))\delta({\rm Im}(z))$, 
we find
\be
\rho(\lambda) = \left .\frac 1\pi \del_{\bar z} G(z)\right |_{z=\lambda} \ .
\label{BCC}
\ee
One of the aims of the present paper is to understand how $G(z) $ develops
nonanalyticities in the complex $z$-plane.
 
Clearly, $G(z)$ is an analytic function of $z$ outside the support of the 
spectrum of the Dirac operator. All singularities must be inside the support.
 In general, $G(z)$ will have cuts in the complex plane
localized within the support of the spectrum. Because of (\ref{BCC}), 
the resolvent
will be a function of both $z$ and $\bar z$ on the support of the spectrum.

The resolvent can be interpreted naturally in terms of a two-dimensional
electrostatic problem \cite{everybody}. 
The electric field is given by the real 
and imaginary parts of the resolvent and the 
spectral density can be interpreted as the charge density.
Eq.\, (\ref{BCC}) is then the two-dimensional analogue of Gauss's law:
\be
\rho(\lambda_r, \lambda_i) = \left .\frac 1{2\pi}\left ( \del_{x} {\rm Re} 
G(z,\bar z) +
\del_{y} {\rm Im} G(z,\bar z)\right ) \right|_{(x,y) =(\lambda_r,\lambda_i)} \ .
\ee

\section{Simple Models}
\subsection{Eigenvalues distributed homogeneously in the complex unit circle}
To illustrate the appearance of nonanalyticities in a function that
shows only an explicit dependence on $z$, we consider the resolvent of 
eigenvalues distributed uniformly inside the complex unit
circle. Using the electrostatic analogy we can conclude 
without calculation that
\be
G(z) = \theta(|z| -1) \frac 1 z + \theta(1-|z|) \bar z \ .
\ee

Let us see how we can understand this result using the average over the spectral
density
\be
G(z) = \int_{\rm unit\,\,\, circle} d^2\lambda \frac 1{z-\lambda} \ .
\ee
In polar coordinates, this integral can be written as
\be
G(z) = \int_0^1 r dr \int_C \frac {du}{-i} \frac 1{(zu-r)} \ ,
\ee
where the contour integral is along the complex unit circle. The integrand
has a simple pole at $u = r/z$. If $|z| < 1$, we obtain a contribution only 
for $r < |z|$ which results in the nonanalyticity mentioned above.

\subsection{Solution of the $2\times 2$ matrix problem}
In this section we study the resolvent for the case when $D$ is a $2\times 2$
matrix. The matrix $z-D$ can be inverted and leads to the average
resolvent 
\be
G(z) = \frac {z}{\pi} \int_0^\infty t dt e^{-t^2} \int_0^{2\pi} d\phi
\frac 1{z^2+t^2-\mu^2 - i\mu t(e^{i\phi} + e^{-i\phi})} \ .
\label{res2}
\ee
We write the angular integral as a contour integral
along the complex unit circle
\be
G(z) = \frac z {\pi i}\int_0^\infty tdt e^{-t^2} \int_C \frac {du}
{u(z^2+t^2-\mu^2) - i\mu t(1+u^2)} \ .
\ee
The poles of the integrand are given by the roots $u_1$ and $u_2$ of
$u(z^2-t^2+\mu^2) - i\mu t(1+u^2) = 0$,
\be 
u_{1,2} = -i \frac{z^2+t^2-\mu^2}{2\mu t} \pm i \left ( 1 + 
\frac{(z^2+t^2-\mu^2)^2}{4\mu^2 t^2} \right )^{1/2} \ ,
\ee
where $u_1$ is the root with the plus sign.
Clearly, the product of the
two roots is $1$. Therefore nonanalyticities can only appear if
the roots cross the unit circle, i.e., for $|u_{1,2}| = 1$.
From the expression for the roots we can see that this happens if
\be
\frac {z^2 +t^2 -\mu^2}{2\mu t} &&\quad {\rm is \,\,\, purely \,\,\, 
imaginary,\,\,\, and}\nonumber\\
\left |\frac {z^2 +t^2 -\mu^2}{2\mu t}\right | &&< 1 \ . 
\ee
With the aid of the real quantity $t_c$ defined below, these conditions 
are equivalent to 
\be
t^2 &=& t^2_c \equiv \mu^2 - \frac 12(z^2 +\bar z^2), \nonumber\\
|{\rm Re}\, z| &\le& \mu \ .
\ee

At the critical point the two roots interchange resulting in the resolvent
\be
G(z) = 2z\int_0^\infty t e^{-t^2} 
\frac 1{\mu t} \frac 1{u_1-u_2} 
-4z\theta(\mu-|{\rm Re}\, z| )  
\int_0^{t_c} t e^{-t^2} \frac{1}{\mu t} \frac 1{u_1-u_2}  \ .
\ee
The first term is analytic in $z$; the second term, (through $t_c$)
depends both on $z$ and $\bar z$.
The spectral density is then given by
\be
\rho(\lambda_r,\lambda_i) = 
\left . \frac 1 \pi \del_{\bar z} G(z) \right |_{z=\lambda_r + i \lambda_i}
 = \theta(\mu-|\lambda_r|) \frac 1\pi
\frac{(\lambda_r^2 + \lambda_i^2) 
e^{-(\mu^2-\lambda_r^2+\lambda_i^2)}}{\sqrt{\mu^2-\lambda_r^2}\sqrt{\mu^2 
+\lambda_i^2}} \ .
\ee

It is straightforward to repeat this calculation for the unquenched case. 
The condensate can be obtained from the partition function which, for one
flavor, is given by
\be
Z(m) =  \int_0^\infty t dt e^{-t^2} \int_0^{2\pi} d\phi \,\det (m-D) \ ,
\label{Z(m)}
\ee
where
\be
\det (m-D) = m^2 +t^2 -\mu^2 -i\mu t(e^{i\phi} + e^{-i\phi}) \ .
\ee
The integrals are elementary resulting in the partition function
\be
Z(m) = \pi(m^2-\mu^2 + 1) \ .
\ee
The corresponding chiral condensate is given by
\be
\Xi(m) = \frac {2 m}{m^2 - \mu^2 + 1} \ .
\ee

For one flavor the resolvent is defined by
\be
G^{N_f = 1}(z) = \frac 1{Z(m)}\int_0^\infty t dt e^{-t^2} \int_0^{2\pi} 
d\phi \, \frac 12 {\rm Tr} \frac 1{z-D}\det (m-D) \ ,
\ee
which can be written as
\be
G^{N_f = 1}(z) = \frac 1{m^2-\mu^2+1}\left[
 z +(m^2- z^2)G^{N_f =0}(z)\right ] \ .
\ee
We see that nonanalyticities are absent for $ z= m$. This was to be 
expected because the singularities of the resolvent are cancelled by the 
zeros of the determinant. A similar cancellation takes place in the Gibbs 
model to be discussed in the next section.

\vskip 1.5cm
\section{The resolvent in the Gibbs model}
\vskip 0.5cm
In this section we analyze the resolvent of the model proposed by Gibbs 
\cite{Gibbs}. 
In this model the Dirac operator is defined by
\be
D^G_{kl} + m\delta_{k,l} = -\delta_{k,l+1} e^{-i\theta -\mu} +
\delta_{k,l-1} e^{i\theta +\mu} +  m\delta_{k,l} \ ,
\ee
where the indices are modulo $N$. Anti-periodic boundary conditions
result in an extra minus sign for the matrix elements $D^G_{1n}$
and $D^G_{n1}$. 
The partition function is obtained by averaging over $\theta$,
\be
Z= \int d\theta {\det}^{N_f}( D^G + m) \ .
\ee
The eigenvalues of $D^G$ are given by 
\be
\lambda_k = e^{\frac{2\pi i (k+1/2)}{N} +i\theta +\mu} -
e^{-\frac{2\pi i (k+1/2)}{N} -i\theta -\mu} \ ,
\ee
where $k = 1, \cdots, N$. Geometrically they fall on an ellipse
in the complex plane with semiaxis $e^\mu - e^{-\mu}$ in the real direction and
semiaxis $e^\mu + e^{-\mu}$ in the imaginary direction.
The fermion determinant follows from the solution
of a recursion relation
\be
\det D^G = e^{-N(i\theta +\mu)}+ e^{-N(i\theta +\mu)} 
+ \lambda_2(m)^N + \lambda_1(m)^N  \ ,
\ee
where 
\be
\lambda_{1,2}(z) = \frac z2 \mp \sqrt{1 +\frac {z^2}4} \ .
\ee
The resolvent, defined by
\be
G(z) = \frac 1Z\int d\theta {\det}^{N_f} 
D^G \frac 1N \sum_k \frac 1{z-\lambda_k} \ ,
\ee
will be evaluated for $N_f = 0$ and $N_f = 1$. (Here, $Z$ is the partition
 function)

The integral over $\theta$ can be performed conveniently by contour
integration. The phase of the eigenvalues and the chemical
potential can be absorbed in $\theta$ resulting in an integral over a circle
 in the complex plane with radius $e^{\mu}$. For $N_f = 1 $ we find
\be
G(z) =\theta(|\lambda_2(z)| - e^{\mu})
\frac 1{\lambda_2(z) - \lambda_1(z)} \left ( 1 - \frac{2\lambda_1^N(z)
}{\lambda_1^N(m) + \lambda_2^N(m)}\right )~~~~~~~\nonumber \\
 ~~~~~~~+ \theta(e^{\mu}-|\lambda_2(z)|)
\frac 1{\lambda_2(z) - \lambda_1(z)} \frac{\lambda_2^N(z)
-\lambda_1^N(z)}{\lambda_1^N(m) + \lambda_2^N(m)} \ .
\ee
This is valid with the convention that $|\lambda_2(z)| \ge |\lambda_1(z)|$.

Note that $|\lambda_1| = |\lambda_2|= 1$ when $z=is$ is purely imaginary 
and  $s < 2$ and positive. For $s>2$ our convention reads  
\be
\lambda_1 &=& \frac{is}2 - \frac i2 \sqrt{s^2 -4} \ ,\nonumber\\
\lambda_2 &=& \frac{is}2 + \frac i2 \sqrt{s^2 -4} \ .\nonumber\\
\ee
The condition $|\lambda_2(z)| = e^\mu$ can be rewritten as
\be
 z &=& e^\mu - e^{-\mu} \qquad {\rm for} \quad z \quad{\rm real} \nonumber \\
 z &=& e^\mu + e^{-\mu} \qquad {\rm for} \quad z \quad{\rm imaginary} \ . 
\ee
This implies that $z$ coincides with the modulus of the eigenvalues on 
the real and imaginary axis, respectively.


In the quenched approximation, we obtain
\be
G(z) = \theta(|\lambda_2(z)|-e^{\mu}) \frac 1{\lambda_2(z) - \lambda_1(z)} \ ,
\ee
which is valid for the same convention, $|\lambda_2(z)| \ge |\lambda_1(z)|$.

We find that quenched and unquenched results agree in the thermodynamic limit 
if $|z|$ is larger than the modulus of any of the eigenvalues. 
For $N\rightarrow \infty$ and  real $z=r$ inside 
the ellipse of eigenvalues, the unquenched resolvent diverges 
for $m < r < e^\mu - e^{-\mu}$ and is zero for $ r<m$.
The quenched result is zero. Quenching works if $m$ is outside the 
ellipse of eigenvalues. For imaginary $z=is$, quenching succeeds for 
$s < \sqrt{m^2 +4}$ but fails in the region $\sqrt{m^2+4} < s < e^\mu + 
e^{-\mu}$. For $m > e^\mu - e^{-\mu}$ quenching works for all values of $s$.

\begin{center}
\begin{figure}[!ht]
\centering\includegraphics[width=100mm,angle=-90]{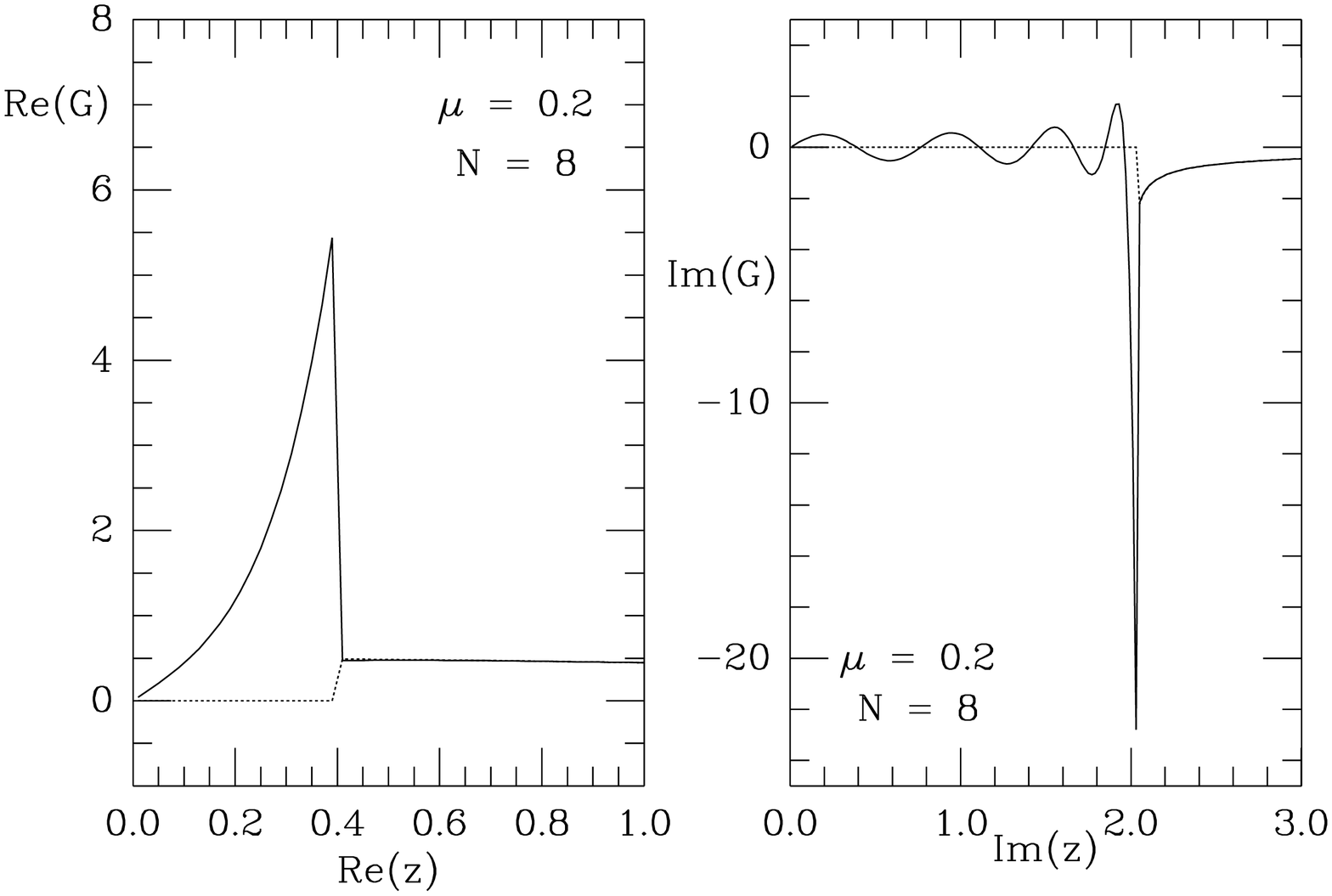}
\begin{center}
\begin{minipage}{13cm}
\baselineskip=12pt
{\begin{small}
Fig. 1. The real (left) and the imaginary (right) parts of the resolvent 
in the Gibbs model on the real and imaginary axis, respectively. Results are 
shown for $\mu = 0.2$ and $n= 8$. The full curve represents the result 
for one flavor and the dashed curve for zero flavors.\end{small}}
\end{minipage}
\end{center}
\end{figure}
\end{center}

  In Fig. 1 we illustrate the above quenched and unquenched results for the 
case $m = 0$, $\mu = 0.2$ and $N= 8$. 
We show the resolvent for $z$ real (left) and $z$
imaginary (right). This figure clearly shows the region where quenching
fails.

It is instructive to consider the thermodynamic limit of $G(z)$ for
$z \rightarrow m$ in the unquenched case. We expect that this limit exists
because the singularities of the resolvent are cancelled by the zeros of
the determinant. Indeed, for $z\rightarrow m$ we find
\be
G(z) = \theta(|\lambda_2(z)| -e^{\mu})\frac 1{\lambda_2(m) - \lambda_1(m)}
+ \theta(e^{\mu}-|\lambda_2(z)| )
\frac 1{\lambda_2(m) - \lambda_1(m)} \left ( \frac{\lambda_2(z)}
{\lambda_2(m)}\right )^N \ ,
\ee 
which is finite for $z=m$. The result for $G(z)$ is independent
of $z$ and identical to $\frac 1N \del_m \log Z$ in the thermodynamic 
limit.

\section{Random matrix model}
In this section we study the partition function
\be
Z(m,\mu) = \int D W P(W) 
det^{N_f} \left ( \begin{array}{cc} m & iW + \mu\\
iW^\dagger +\mu & m \end{array} \right) \ ,
\label{zmu}
\ee
where $W$ is an arbitrary complex $n\times n$ matrix and $DW$ the Haar measure.
The probability distribution $P(W)$ is given by
\be
P(W) = \exp (-n {\Sigma^2} {\rm Tr} WW^\dagger) \ .
\ee
The fermion determinant in (\ref{zmu})
can be written as a Grassmann integral.
\be
Z(m, \mu) =
\int {\cal D}W  {\cal D} \psi^* {\cal D} \psi
\exp\left[- i\sum_{k=1}^{N_f}\psi^{k\,*} \left (
\begin{array}{cc}  m & iW + \mu \\i W^\dagger +\mu& { m}
\end{array} \right )
\psi^k \right]  \exp[-{n\Sigma^2} \,{\rm Tr}\, WW^\dagger] \ \ .\nonumber 
\\
\label{ranpart}
\ee
This enables us to perform the $W$ integration
\be
Z(m, \mu) = \int {\cal D} \psi^* {\cal D}\psi \exp \left[ 
-\frac 1{n\Sigma^2 } \psi^{f\,*}_{L\, k}
                     \psi^{f}_{R\, i} 
                     \psi^{g\,*}_{R\,i}
                     \psi^{g}_{L\,k } \right. + 
           { m}\left(\psi^{f\,*}_{R\, i}
                     \psi^{f}_{R\, i}+
                     \psi^{f\,*}_{L\,k}
                     \psi^{g}_{L\, k}\right)  
                   \nonumber \\
                 +  \left. \mu \left(\psi^{f\,*}_{R\, i} 
                     \psi^{f}_{L\, i}
                    +\psi^{f\,*}_{L\,k}
                     \psi^{f}_{R\, k}\right) \right] \ .
\ee
The four-fermion terms can be
written as the difference of two squares. Each square can be linearized by
a Hubbard-Stratonovitch transformation according to
\be
\exp(-A Q^2) \sim \int d\sigma\exp(-\frac{\sigma^2}{4A} - iQ \sigma) \ \ .
\label{Hubbard}
\ee
Thus, the fermionic integrals can be performed, and the
partition function can be written as a single integral over the 
complex $N_f\times N_f$ matrix, $\sigma$,
\be
Z(m, \mu) = \int {\cal D} \sigma \exp [-{n\Sigma^2} {\rm Tr}
\sigma \sigma^\dagger] {\det}^{n}
\left ( \begin{array}{cc}  \sigma+{m} & \mu \\  \mu &\sigma^\dagger +{
m}
\end{array} \right ) \ .
\label{apart2}
\ee

The condensate is given by
\be
\langle \bar q q \rangle
 = \frac 1{2n N_f} \del_{m} \log Z \equiv G(m)\ \ .
\ee
For $n \rightarrow \infty$, it can be evaluated with the aid
of a saddle point approximation. The saddle point equations are given by
\be
-{ \Sigma^2} \sigma+ (\sigma+m)\left( (\sigma^\dagger +m)
(\sigma+m) -\mu^2 \right
)^{-1} = 0 \ \ \ ,
\label{spe}
\ee
and an equation with $\sigma$ and $\sigma^\dagger$ interchanged. The solutions
are proportional to the identity with diagonal elements given (for $\Sigma 
=1$) by 
\be
\sigma^* &=&\sigma,\nonumber \\
\sigma(m+\sigma)^2 -\mu^2 \sigma &=& m+ \sigma \ .
\label{cubic}
\ee
This equation was first derived \cite{JV} for the finite temperature version
of the model \cite{JV,Tilo,Stephanov1}. 
(obtained from (\ref{apart2}) by the substitution 
$\mu\rightarrow -iT$).
The condensate is then given by
\be
\langle \bar q q \rangle = G(m) = \sigma \ .
\ee
The quark number density defined in (\ref{quarkdens}) can be related to
$\sigma$,
\be
n_q = \frac {-\mu \sigma}{\sigma + m}
\label{nqsig}
\ee
For $m = 0$ it is possible to have a discontinuity in 
$\langle \bar q q\rangle$ but not in $n_q$. Notice that in the restored phase 
with $\sigma= 0$ this expression has to be treated with care. In fact, from
the saddle point equation we
obtain $n_q = 1/\mu$ for $\sigma = 0$, and  $n_q = -\mu$ for $\sigma \ne 0$. In 
general, we expect that a discontinuity in the 
chiral condensate is accompanied by
a discontinuity in the quark number density.

Two solutions of  (\ref{cubic}) coincide if the discriminant of the cubic
equation, 
\be
D_3 = \frac 1{27}\left( m^4 \mu^2 - m^2(2\mu^4 - 5\mu^2 -\frac14) 
+(1+\mu^2)^3\right) \ ,
\label{discriminant}
\ee
vanishes.

In the finite temperature version of the present model \cite{JV}, 
we evaluated the
average resolvent for $z$ on the positive real  axis and obtained
the resolvent for ${\rm Re}(z) > 0$ through analytical continuation. 
This is justified because 
the resolvent is analytic in this part of the complex plane.
The average resolvent was calculated with two methods \cite{sener}: 
(i) the super-symmetric
method, and (ii) a self-consistent equation for the resolvent for 
$n \rightarrow \infty$. The first method is much more intricate with 
regards to analyticity and convergence than the second. For the moment, we 
discuss only the second method. 
The self-consistent equation was obtained from a series 
expansion of the resolvent $G(z)$ in powers of $z^{-1}$. 
This is justified if
$|z|$ is larger than the largest eigenvalue. In our present model with a 
chemical potential, this is still true. It should also be clear that the 
resolvent is analytic in $z$ everywhere outside of the support of the
eigenvalue spectrum of the Dirac operator.  In other words, the singularities
of $G(z)$ should be inside the support of the spectrum of the Dirac operator.
However, for $z$ inside the domain of eigenvalues, the series 
expansion is not justified. Indeed, the resolvent in this region is 
not given by the cubic equation (\ref{cubic}), as was shown by Stephanov 
\cite{Stephanov2}. After replacing $T\rightarrow i\mu$ and making an 
appropriate rotation in the complex plane (the Dirac operator in 
\cite{JV,sener} differs by a factor $i$ from the Dirac operator in the 
present work), the self-consistent cubic equation in \cite{JV,sener} is 
seen to be identical to the cubic equation (\ref{cubic}).

\subsection{Unquenched partition function}
In this section we analyze the partition function (\ref{apart2}) for one 
flavor.
In units where $\Sigma = 1$, it is given by
\be
Z(m,\mu) = \int d\sigma d\sigma^* e^{-n\sigma^2} (\sigma \sigma^* + 
m(\sigma+\sigma^*) + m^2 -\mu^2)^n \ .
\label{zoneflavor}
\ee
After shifting the real part of $\sigma$ by $m$, we 
use polar coordinates
as new integration variables. If we notice that 
the angular integral represents a modified Bessel function, the partition
function can be written as
\be
Z(m,\mu) = \pi e^{-nm^2}\int_0^\infty du 
(u-\mu^2)^n I_0(2mn\sqrt u) e^{-nu} \ .
\label{zi0}
\ee
In the thermodynamic limit, this partition function can be evaluated
by a saddle point approximation. Using the asymptotic form of 
$I_0(z) \sim e^z/\sqrt{2\pi z}$, the saddle point equation reads
\be
\frac 1{u-\mu^2} = 1- \frac m{\sqrt u} \ .
\label{saddleu}
\ee
A phase transition takes place 
at the points where 
$| Z_{u=u_b}|= | Z_{u=u_r} |$, with $u_b$ and $u_r$ two solutions of
the saddle-point equation (\ref{saddleu}). 
This condition can be rewritten as
\be
|(u_b-\mu^2) e^{2m\sqrt u_b -u_b}|= 
|(\mu^2-u_r) e^{2m\sqrt u_r -u_r}| \ .
\label{critical} 
\ee
The selection of the dominant saddle-points requires a detailed analysis
of the partition function (see below).
This will lead to some restrictions on the range of $\mu$ and $m$ 
determined by the zeros of the discriminant (\ref{discriminant}). 
Below we will show that (\ref{critical}) determines the
location of the zeros of the partition function. The  
additional limitations on the range of the critical
values of $m$ and $\mu$ immediately follow from such analysis.

In general, the solution of these equations is cumbersome. However,
for $m \rightarrow  0$ we find that $u_r = 0$ and 
$u_b = 1+\mu^2$. This leads to
the critical curve
\be
{\rm Re}\left [ 1+\mu^2 +\log \mu^2 \right ] = 0 \ .
\label{crmu}
\ee
For real $\mu$, the solution is given by $\mu_c = 0.527\dots$; for purely 
imaginary $\mu$, we find $\mu_c = i$. Moreover, from a detailed
saddle-point analysis it can be shown that the partition function
only shows a discontinuity for those values of $\mu$ that in addition
to (\ref{crmu}) also  satisfy  $|\mu| < 1$. 

We wish to analyze the zeros of the partition function (\ref{zoneflavor}) 
in the complex $m$ plane and the complex $\mu$ plane. 
In order to obtain the polynomial in $m$ and $\mu$, we 
expand the binomial in (\ref{zi0}) and perform the integration over $u$. 
The result can be expressed using confluent hypergeometric functions
\be
Z(m,\mu) = \frac {\pi n! e^{-nm^2}}{n^{n+1}}\sum_{k=0}^n \frac{(-\mu^2 n)^k} 
{k!} ~_1F_1((n-k+1),1,m^2n) \ .
\ee
According to the Kummer identity we have
\be
e^{-nm^2}  {}_1F_1((n-k+1),1,m^2n) = {}_1F_1(-(n-k),1,-m^2n)
\ee
which is a polynomial of order $n-k$ in $-m^2 n$. From the standard  
representation of this Kummer function we obtain
\be
Z(m,\mu) = \frac{\pi n!}{n^{n+1}}\sum_{k=0}^n\sum_{l=0}^{n-k} 
 \frac 1{l!k!} \left ( \begin{array}{c} n-k \\ l \end{array} \right )
(-\mu^2 n)^k (m^2n)^l  \ .
\label{finiteseries}
\ee
Although the sums in (\ref{finiteseries}) are finite, the alternating nature
of the series and the degree of cancellation
make numerical evaluation
difficult. 

For real $\mu$ the phase structure of the partition function (\ref{zi0})
can be clarified by the decomposition 
\be
Z(m,\mu) = Z^{\rm broken}(m, \mu) + Z^{\rm restored}(m, \mu) \ ,
\ee
where
\be
Z^{\rm broken}(m, \mu)&=& 
\pi e^{-nm^2}\int_{\mu^2}^\infty du 
(u-\mu^2)^n I_0(2mn\sqrt u) e^{-nu} \nonumber\\
Z^{\rm restored}(m, \mu)&=& 
\pi e^{-nm^2}\int_{0}^{\mu^2} du 
(\mu^2-u)^n I_0(2mn\sqrt u) e^{-nu} \ .
\label{zsplit}
\ee
From a graphical representation of both sides of 
the saddle point equation (\ref{saddleu})
it is clear that one solution is in the interval $[0,\mu^2]$, which we call 
$u_r$, another 
is in the interval $[\mu^2, \infty)$, which we will call $u_b$ 
and another has $\sqrt u < 0$. 
A phase transition takes place at the point where  
$| Z_{u=u_b}|= | Z_{u=u_r} |$. 
For $m=0$ 
the thermodynamic limit of  the partition function is then given by
\be
Z(m=0, \mu)&=& \theta(\mu_c-\mu) \frac{\pi^{3/2}}{\sqrt n} e^{-n(1+\mu^2)}
+\theta(\mu-\mu_c) \frac{\pi^{3/2}}{\sqrt n} \mu^{2n} \ .
\label{zm=0}
\ee

The partition function can be calculated in a much more effective 
way if we rewrite it as a sum of positive definite terms. For the 
broken part, we expand $I_0$ in a power series and change integration 
variables according to $u \rightarrow \mu^2(1+u)$. The integrals can 
be performed following expansion of the binomial and result in
\be
Z^{\rm broken}(m, \mu) =
\frac{\pi e^{-n(m^2+\mu^2)}}{n^{n+1}} \sum_{l=0}^\infty 
\sum_{s=0}^l \frac {(n+s)!}{l! s!(l-s)!} (m^2 n)^l (\mu^2 n)^{l-s} \ .
\ee
For the restored part, we also expand the modified Bessel function 
but change integration variables according to $u \rightarrow \mu^2(1-u)$. 
The integral of each term in the expansion of $\exp(n\mu^2 u)$ is a beta 
function. Collecting powers and factorials we obtain
\be 
Z^{\rm restored}(m, \mu) =
\pi \mu^{2(n+1)} e^{-n(m^2+\mu^2)} \sum_{l=0}^\infty 
\sum_{s=0}^\infty \frac {(n+s)!}{l! s!(n+s+l+1)!} (m^2 n)^l (\mu^2 n)^{l+s} \ .
\ee
This expression has been used for a numerical study of the partition function.

\subsection{Yang-Lee zeros}
 In this section we evaluate the Yang-Lee zeros of the partition function
(\ref{finiteseries}) and show that their location is consistent with a
saddle-point analysis of the partition function. 
First, we consider the partition function as a polynomial
in $m$.  Results for $\mu = 0$, $\mu = 0.5$, and $\mu = 0.6$ are shown
in the left column of Fig.\,2. 
We have calculated the zeros for different values of $n$, 
e.g., $n=48$,
$n=96$, and $n=192$.  The results for $n=192$ are represented by the points
in the figure.  Of course, the exact location of the zeros is extremely
sensitive to numerical round-off errors. Thus, the present results were
obtained with the help of a multi-precision package \cite{bailey}.
Typically, we performed our computations with 100-500 digits accuracy.

\begin{center}
\begin{figure}[!ht]
\centering\includegraphics[width=70mm]{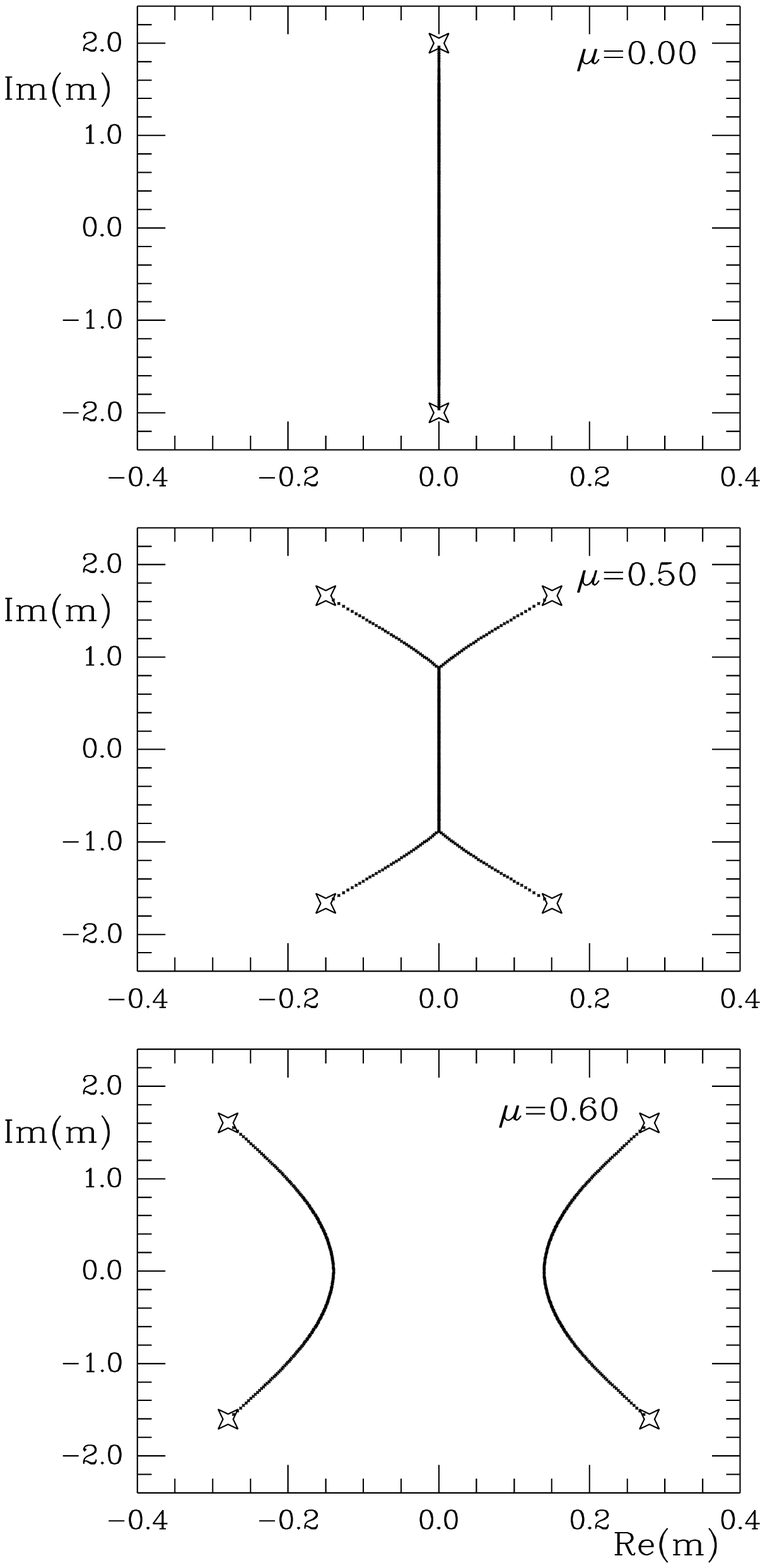}
\includegraphics[width=70mm]{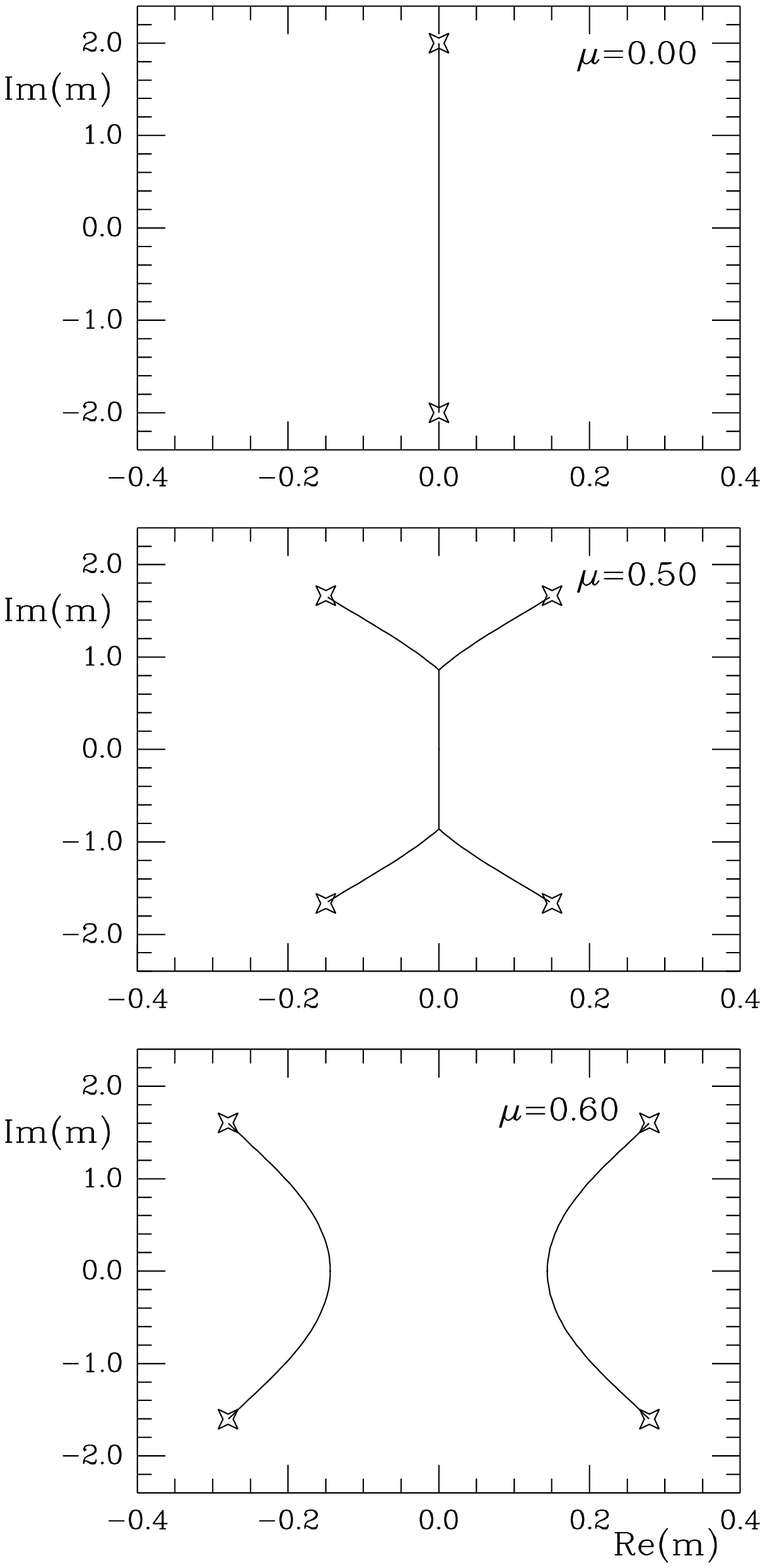}
\begin{center}
\begin{minipage}{13cm}
\baselineskip=12pt
{\begin{small}
Fig.\,2. The zeros of the partition function in the complex $m$ 
plane (left), and the curves where the mean field result for the resolvent 
shows a discontinuity (right). Results are given
for $\mu =0$ (upper), $\mu = 0.50$ (middle) and $\mu = 0.60$ (lower)
for $n = 192$.  Zeros of the discriminant of the cubic equation
(\ref{cubic}) are denoted by stars.\end{small}}
\end{minipage}
\end{center}
\end{figure}
\end{center}

The zeros fall on a curve and are regularly spaced\footnote{If the numerical
accuracy is not sufficient, one typically observes that the line of zeros ends
in a circle.}.
It is clear from the saddle point analysis of the partition function 
that a single condition is imposed on the the complex variable $m$ by the 
condition that the free energy of two different saddle point solutions 
coincides.
This explains the fact that, in the thermodynamic limit, the zeros 
are located on a curve in the complex plane. (A similar argument has 
been given for the Ising model \cite{Shrock}.) 
If we increase the order of the polynomial by a factor of
$2$, we find that half of the zeros are close to those of the lower-order
polynomial.  
The other half of the zeros are roughly mid-way between adjacent zeros
of the lower-order polynomial.  This leads us to the conclusion that the
zeros become dense and lead to a cut in the complex $m$ plane in the
thermodynamic limit.

\begin{center}
\begin{figure}[!ht]
\centering\includegraphics[width=70mm]{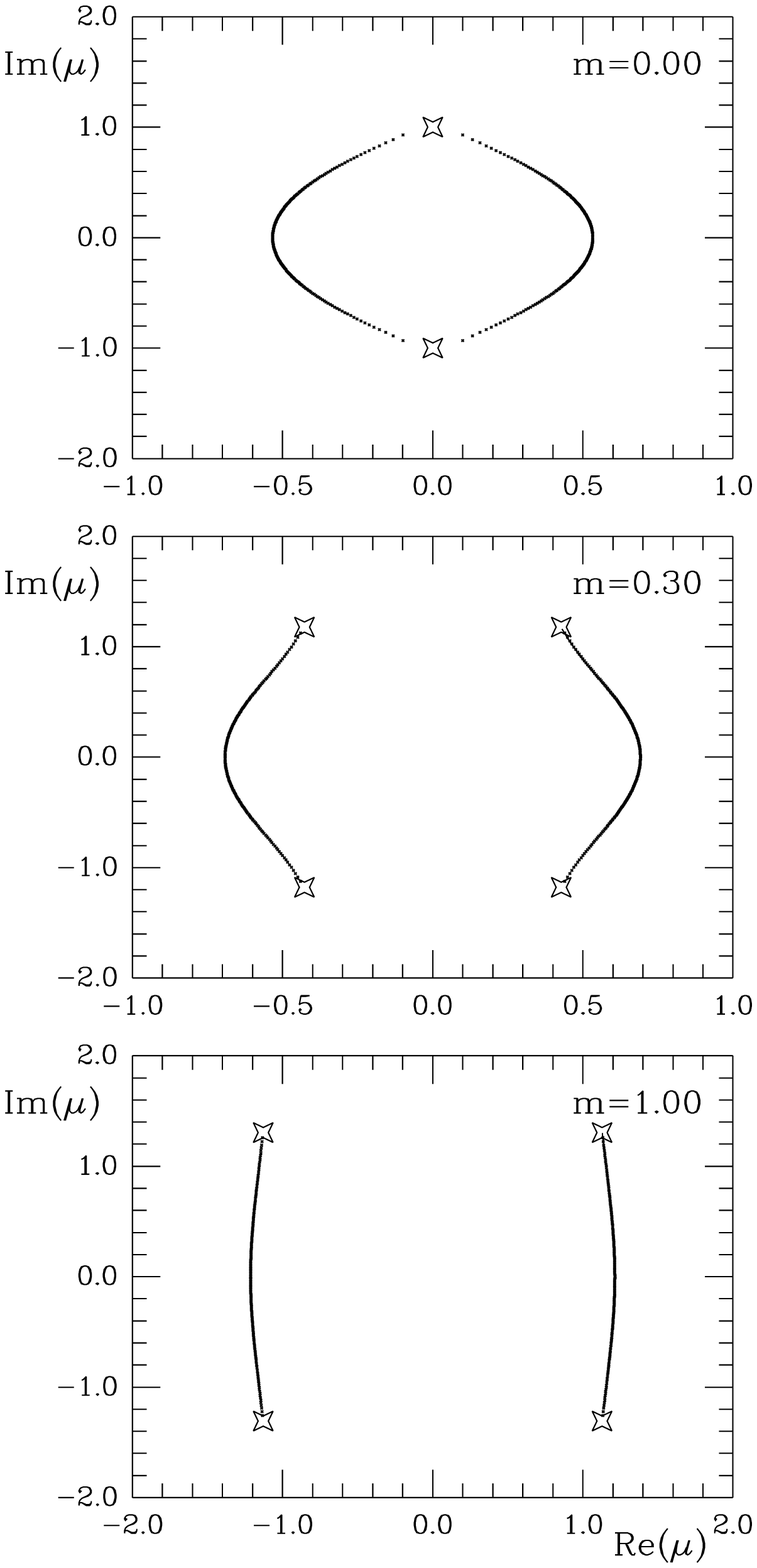}
\includegraphics[width=70mm]{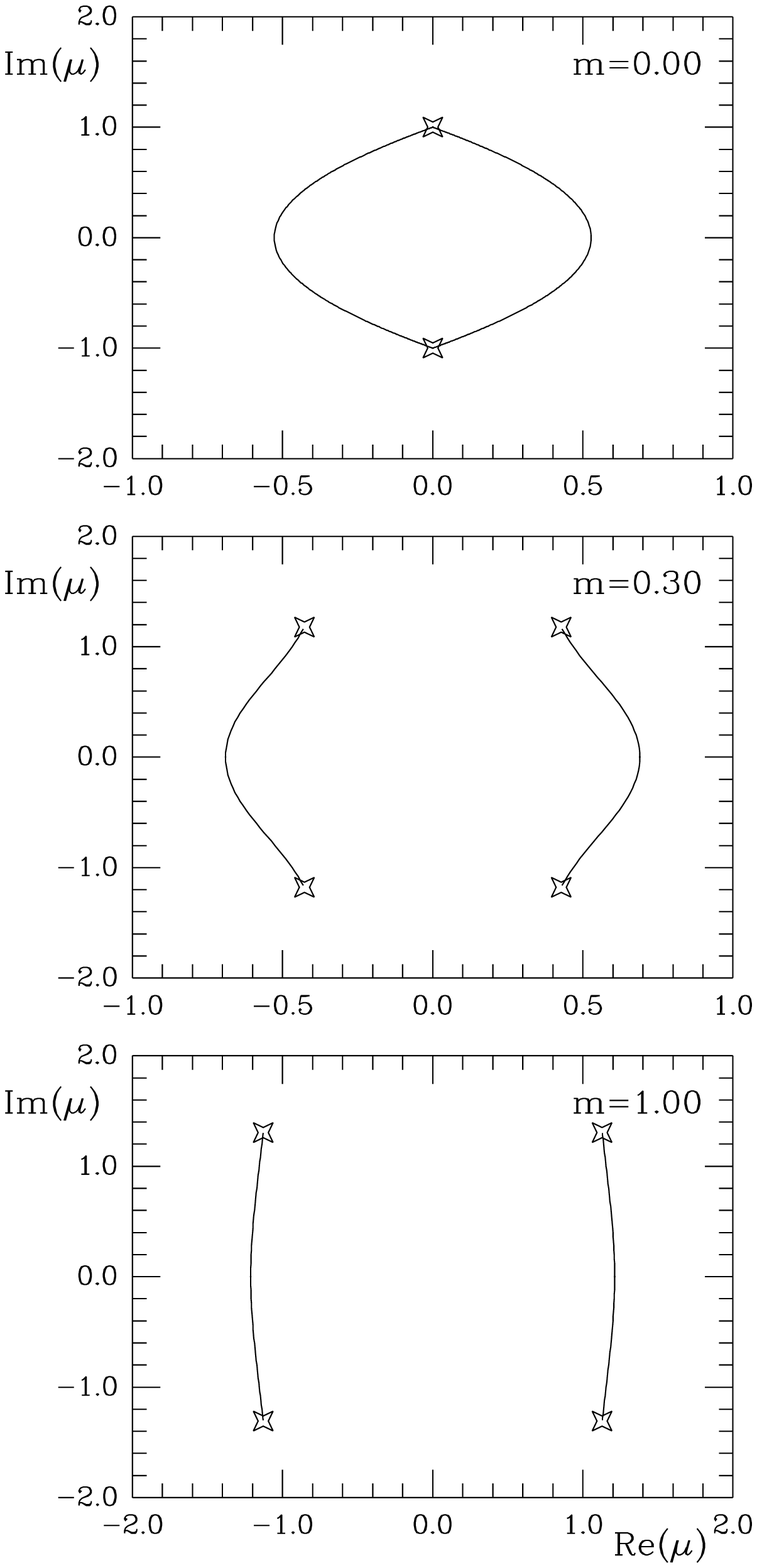}
\begin{center}
\begin{minipage}{13cm}
\baselineskip=12pt
{\begin{small}
Fig.\,3. The zeros of the partition function in the complex $\mu$ 
plane (left) and the location of points where the mean field result for 
the chiral condensate shows a discontinuity (right). 
Results are given for
 for $m =0$ (upper), $m = 0.30$ (middle), and $m = 1.0$ (lower) for
$n = 192$. The stars represent 
zeros of the discriminant of the cubic equation (\ref{cubic}).
Note that the scale on the $x$-axis of 
the lower figure is different.\end{small}}
\end{minipage}
\end{center}
\end{figure}
\end{center}

The stars in Fig.\,2 represent the points at which the discriminant
(\ref{discriminant}) of the cubic saddle point equation vanishes. These
points coincide with the endpoints of the line of zeros.  
In the right column of Fig. 2 we show the curves in the complex $m$-plane 
across which there is a transition in the dominant solution of the 
third-order equation (\ref{cubic}).
The analytical result for this curve is given in (\ref{critical}).
A schematic picture of the latter was also shown in 
\cite{stephanovlat}.

In  the left column of Fig.\,3, we show the zeros of the partition function
in the complex $\mu$ plane for n=192 and masses
$m = 0$, $ m = 0.3$, and $m= 1.0$.
We have evaluated the zeros for $n=48$ and $n=96$. Here, too, 
all calculations were performed with 100-500 digits accuracy. The zeros
are regularly spaced, and their density increases homogeneously with $n$. In 
the thermodynamic limit, we therefore expect that they join into a cut. In all
cases, the line of zeros terminates at a zero of the discriminant
(\ref{discriminant}), which is denoted by a star. In right column of Fig. 3, 
we show the curve across which the saddle point result for the chiral
condensate shows a discontinuity. Below (\ref{nqsig}) it was argued that 
this will lead to a discontinuity in the quark number density as well. Notice
that for $m = 0$ the chiral condensate jumps from a finite value to zero.
The analytical result follows from the   
transcendental equation (\ref{critical})
For $m= 0$, chiral symmetry is broken in the region
enclosed by this curve and is restored outside.
In this case, the density of zeros approaches zero near $\mu = i$.  This is not
surprising since, at this point, the phase transition 
changes from  first to second order.

\section{Quark number density}
In this section we study the effect of the fermion determinant
and eigenvalue correlations on the quark number density defined in
(\ref{quarkdens}). In the quenched approximation the quark number 
density is given by
\be
n_q =  \frac 1N {\rm Tr} 
\left (
\begin{array}{cc} iW+ \mu & m  \\m& iW^\dagger +\mu 
\end{array}\right )^{-1} \ ,
\ee
where $W$ is distributed according to (\ref{prob}). For  $m$ equal to 
zero, the quark number density follows from the spectral
density of the ensemble of matrices
\be
\left (
\begin{array}{cc}  iW & 0\\ 0&iW^\dagger   \end{array}\right ) \ .
\ee
This ensemble was analyzed by Ginibre \cite{Gin}. 
An arbitrary complex matrix can be diagonalized by  a similarity transformation
\be
W= X \Lambda X^{-1} \ ,
\ee
where $\Lambda$ is a complex diagonal matrix. 
In this case, the Gaussian probability distribution depends both on 
$\Lambda$ and $X$. Thus, the problem of 
determining the joint probability distribution of the eigenvalues
is not solved by simply choosing the eigenvalues and eigenvectors as
new integration variables. However, using a remarkable precursor of 
an Itzykson-Zuber integral, Ginibre was able to solve the problem 
analytically. In the thermodynamic limit, the eigenvalues of $W$ (and 
therefore of $W^\dagger$, $iW$ and $iW^\dagger$) are distributed 
uniformly in the complex unit circle. As we will show below, 
results in a nonzero quark number density for arbitrary small $\mu$. 
In terms of the eigenvalues, the baryon number density is given by
\be
n_q = \frac 1N \sum_k \left [\frac 1{\lambda_k + \mu}+ 
\frac 1{\lambda_k^* + \mu} \right ] \ .
\ee
For eigenvalues distributed uniformly inside the unit circle, the baryon 
number density can be evaluated in the same way as the chiral condensate 
in section 2. The result is:
\be
n_q = \theta(1-\mu) \mu + \theta(\mu-1) \frac 1\mu \ .
\ee
For the unquenched case, the baryon number density in the broken phase
is quite different with 
\be
n_q = \theta(\mu_{\rm c}-\mu)(-\mu) + \theta(\mu_{\rm c} -\mu) \frac 1\mu. 
\label{nqunquenched}
\ee

Finally, we wish to stress that the correlations of the eigenvalues play
an important role in establishing the result (\ref{nqunquenched}). For 
uncorrelated eigenvalues distributed homogeneously inside the complex 
unit circle, the unquenched partition function factorizes into 
\be
Z(\mu) &=&\left [\frac 1\pi \int_D d^2\lambda (\mu^2 
-|\lambda|^2+i\mu(\lambda+\lambda^*))\right]^n
\nonumber \\
          &=& (\mu^2 -\frac 12)^n \ .
\label{zminus}
\ee
The quark number density, which is then given by 
\be
n_q = \frac {\mu}{\mu^2 - \frac 12} \ ,
\ee
shows some of the characteristic features of the unquenched result, 
including a discontinuity at a critical value of $\mu$. In terms of 
the joint eigenvalue distribution of the random matrix ensemble 
(\ref{rmt}), the partition function is given by 
\be
Z(\mu) & =& \frac 1{\cal N}\int \prod_k d\lambda_k d\lambda_k^* 
\prod_k[\mu^2 - |\lambda_k|^2 +i\mu(\lambda_k+\lambda_k^*)] 
\prod_{k<l}|\lambda_k 
-\lambda_l|^2 \exp(-|\lambda|^2) \ ,
\label{zccor}
\ee
where ${\cal N}$ is a normalization constant. The correlations between the
eigenvalues which are mainly induced by the Vandermonde determinant in
(\ref{zccor} leads to a partition function that is completely different
from (\ref{zccor}). (In the thermodynamic limit
(\ref{zccor})  reduces to (\ref{zm=0})).
In particular, the zeros of (\ref{zccor}) are $n$-fold degenerate,
whereas the zeros of
(\ref{zoneflavor}) are located on a closed curve as shown in Fig. 3.

It is equally straightforward to obtain the result with the determinant 
replaced by its absolute value. 
Then, 
\be
Z(\mu) &=&\left [\frac 1\pi \int_D d^2\lambda (\mu^2 
+|\lambda|^2+i\mu(\lambda+\lambda^*))\right]^n
\nonumber \\
          &=& (\mu^2 +\frac 12)^n \ .
\ee
The quark number density, now given by
\be
n_q = \frac {\mu}{\mu^2 + \frac 12} \ ,
\ee
is much closer to the quenched result.

We conclude that quenching has a major effect on the quark number density.
We have also found that
correlations between the eigenvalues play an important role in establishing 
the $\mu$-dependence of the partition function. 
 
\vskip 1.5cm
\noindent\section{Numerical results for the unquenched resolvent}
\vskip 0.5cm
In section 4 we demonstrated that, for $z$ less than some critical 
value, the resolvent of the unquenched Gibbs model diverges in the 
thermodynamic limit. We note that this model is special in the sense 
that its eigenvalues fall on a one-dimensional manifold in the complex plane 
(i.e., an ellipse), which is not the case in lattice QCD simulations 
\cite{everybody}. In this section we study the model (\ref{rmt}), 
which has a genuinely two-dimensional spectral distribution. Unfortunately, 
we have not been able to treat the unquenched case analytically. However, 
it can be attacked numerically by brute force, i.e., the fermion determinant 
is not included in the measure but rather in the observable. 
Results for $G(z)$ on the real and imaginary axes are shown 
in Figs. 4, 5 and 6. In Fig. 4 we show the real part of the 
resolvent for $\mu = 0.2$ and $n=24$ (left) and $n=48$ (right). The mass in
 the determinant is equal to zero. The full line represents unquenched results 
for the resolvent. The analytical quenched results are represented by the 
dotted curves. The almost horizontal dotted curve shows the solution
given by the cubic equation (\ref{cubic}). The dotted curve through zero is 
the Stephanov solution \cite{Stephanov2} which is given in general by 
\be
G(z=x+iy) = \frac 12 \frac x{\mu^2 -x^2} -x - \frac {iy}2 \ .
\ee
Up to corrections of order $1/n$ the quenched numerical result (dashed curve)
is in agreement with the Stephanov solution inside the domain of eigenvalues
($|x|<0.0732$ for $\mu = 0.2$ and $|x| < 0.1504$ for $\mu = 0.3$) 
and follows the solution of the cubic equation outside this domain.
In this region the unquenched resolvent agrees with the quenched resolvent up
to corrections of order $1/n$. The most dramatic result of this calculation is
that the real 
part of the resolvent seems to diverge for $z$ inside the domain of 
eigenvalues. This is in complete agreement with the
observation made for  the Gibbs model.
The slope near zero is strictly proportional to 
$n$. This can be understood analytically from the absence of eigenvalues
in this region. Therefore, the resolvent can be expanded as
\be
G(z) = z \frac 1{2n}\sum_k \frac {-1}{\lambda_k^2} + O(z^3) \ .
\ee
The inverse moment of the eigenvalues follows immediately 
from the proper unquenched
partition function (\ref{zoneflavor})
\be
\frac 12\sum_k \frac {-1}{\lambda_k^2} = \del_{m^2} \log Z = n^2 (1+\mu^2)
\ee
resulting in
\be
G(z) = z n(1+\mu^2) + O(z^3) \ ,
\ee
which is in perfect agreement with the numerical results.

Results for the real part of the resolvent for $n=24$ and $\mu =0.3$ are 
shown in Fig. 6. We find that the maximum increases strongly with $\mu$.
This should be contrasted with the same model at finite temperature 
where the average resolvent is flavor independent.

The imaginary part of the resolvent is shown in Figs.\,5 and 6 
as a function of the imaginary part of $z$. The fluctuations of the 
unquenched result are much stronger in this case. Remarkably,  the 
unquenched result seems to agree with the quenched result (dotted curve) 
for all real $z$. The area of the peak near zero is proportional 
to $1/n$ which suggests that this is a finite size effect. Notice that 
inside the domain of eigenvalues the analytical result (i.e., the 
Stephanov solution) does not depend on $\mu$. Only the 
point where the two solutions intersect is $\mu$-dependent.

\begin{center}
\begin{figure}[!ht]
\centering\includegraphics[width=90mm,angle=-90]{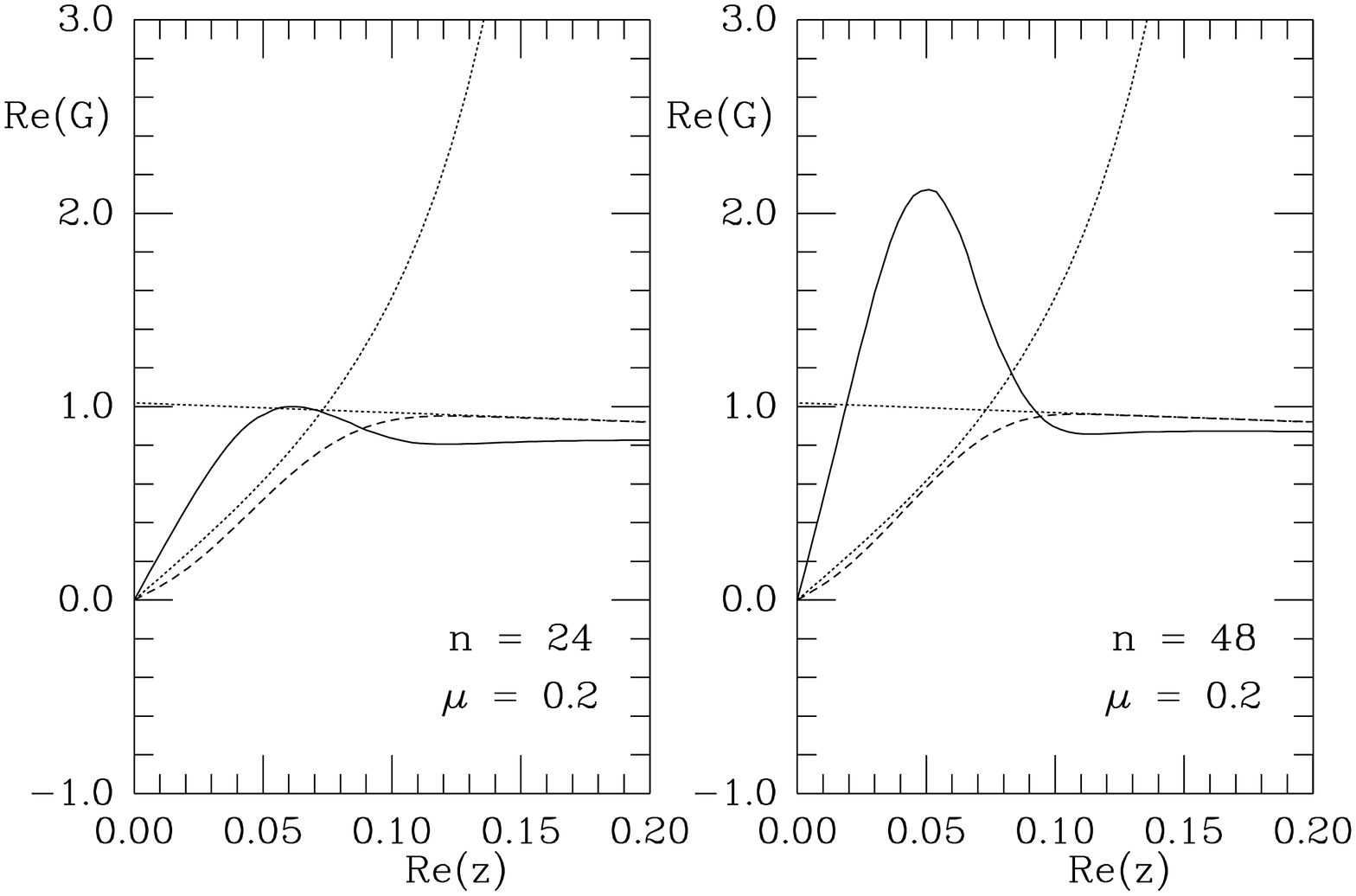}
\begin{center}
\begin{minipage}{13cm}
\baselineskip=12pt
{\begin{small}
Fig. 4. Real part of the resolvent $G(z)$  as a function
of ${\rm Re}(z)$ for ${\rm Im}(z)=~0$.\end{small}}
\end{minipage}
\end{center}
\end{figure}
\end{center}

\begin{center}
\begin{figure}[!ht]
\centering\includegraphics[width=90mm,angle=-90]{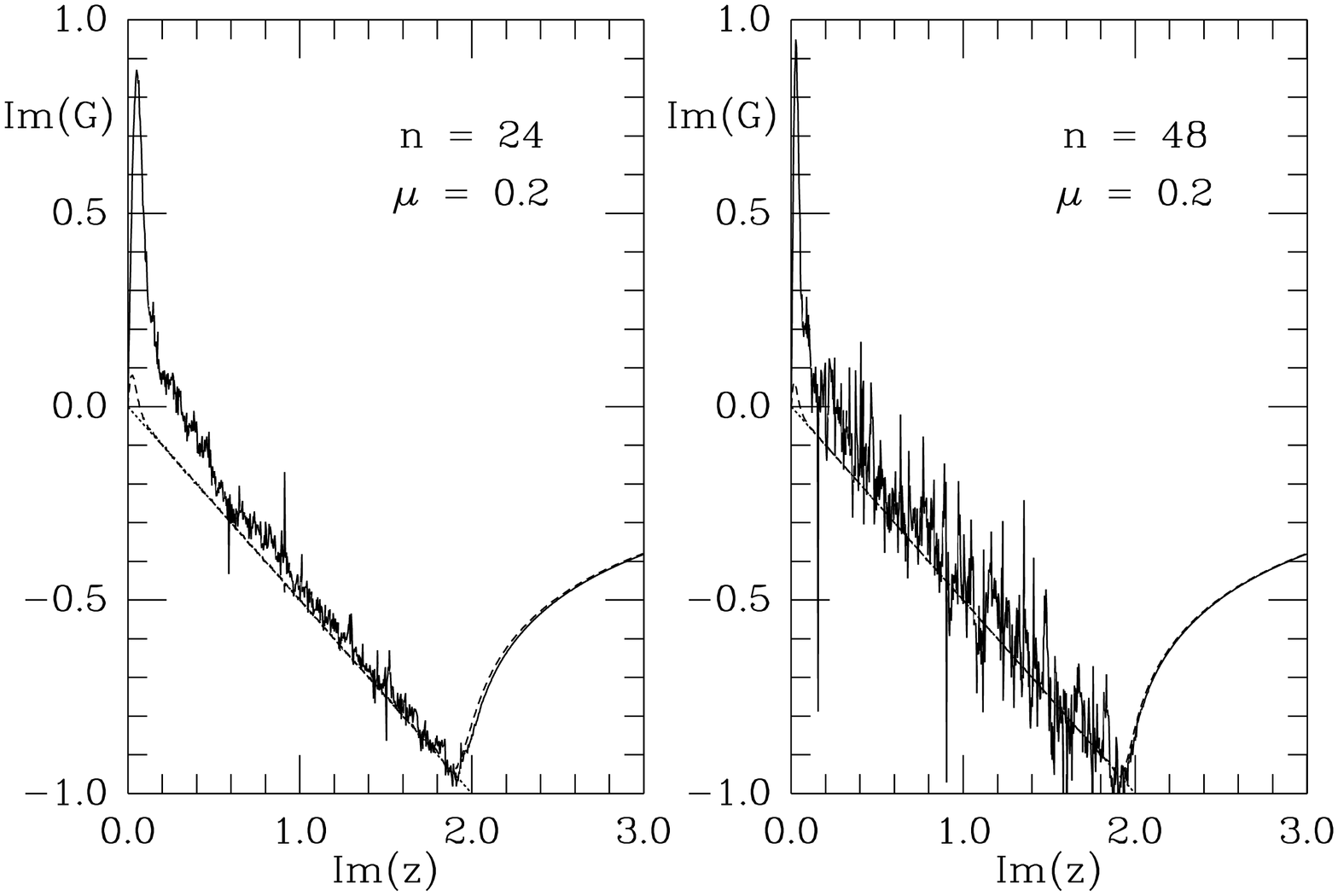}
\begin{center}
\begin{minipage}{13cm}
\baselineskip=12pt
{\begin{small}
Fig. 5. Imaginary part of the resolvent $G(z)$  as a function
of ${\rm Im}(z)$ for ${\rm Re}(z)=~0$.\end{small}}
\end{minipage}
\end{center}
\end{figure}
\end{center}

\begin{center}
\begin{figure}[!ht]
\centering\includegraphics[width=90mm,angle=-90]{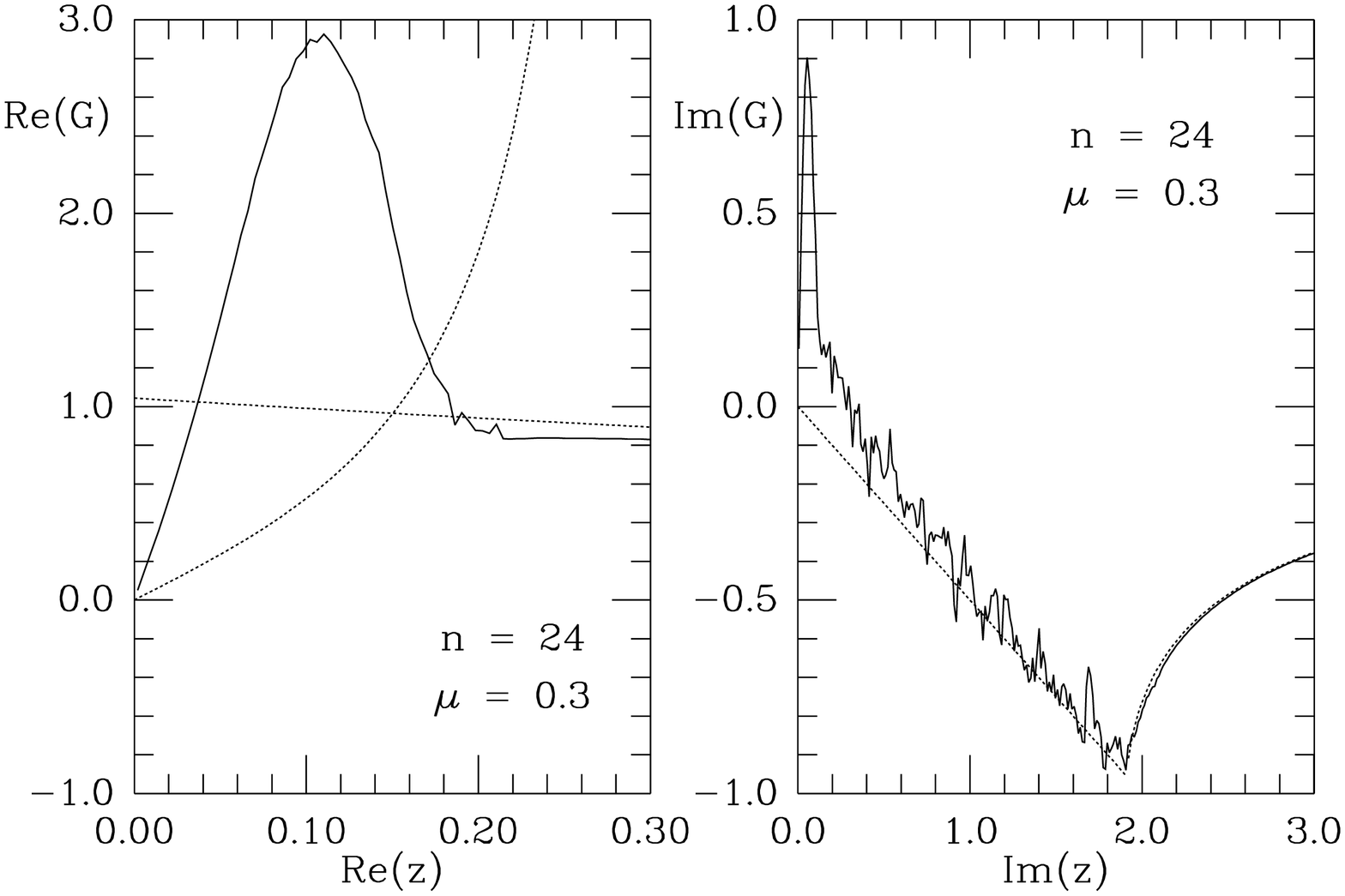}
\begin{center}
\begin{minipage}{13cm}
\baselineskip=12pt
{\begin{small}
Fig. 6. Real and Imaginary parts of the resolvent $G(z)$ as a function
of ${\rm Re}(z)$ and ${\rm Im}(z)$ for ${\rm Im}(z)=0$ and 
${\rm Re}(z)=0$, respectively.\end{small}}
\end{minipage}
\end{center}
\end{figure}
\end{center}

The results for $n=48$ and $\mu = 0.2$ were obtained by averaging over 
one million matrices. Results for $\mu = 0.3$ and $n = 24$ required as many as
three million matrices.
The difficulties in simulating the partition function becomes clearer if
we consider the expectation value of the fermion determinant which is given
by the partition function for one flavor. 
Let us write out explicitly the determinant 
in the definition (\ref{resolvent}) of the unquenched resolvent
\be
G(z) \equiv \frac 1N \left< {\rm Tr} \left( \frac 1{z-D} \right) \right> 
\equiv 
\frac{ \left<\left< {\det}^{N_f}(m-D) {\rm Tr} 
\left( \frac 1{z-D} \right) \right>\right> }
{ \left<\left< {\det}^{N_f}(m-D) \right>\right>} \ ,
\ee
where $\left<\left<{\cdots}\right>\right>$ stands 
for averaging without including
the determinant in the weight. 
The normalization is just the unquenched partition function. 
For one flavor, it was evaluated in (66). In the broken phase it is given by
\be
Z(m=0, \mu) = \frac{\pi^{3/2}}{\sqrt n} e^{-n(1+\mu^2)}
\ee
This exponential suppression makes it prohibitively difficult to 
calculate the unquenched resolvent
via a Monte Carlo method.
Indeed, we find that the convergence of numerical simulations
becomes extremely slow for $\mu^2n > 1$. 
The average resolvent is the sum of terms weighted 
with a typical value of the determinant. Because of the normalization we
expect an enhancement factor
of $\sim e^{n\mu^2}$.  
The ratios of the maximum values of the real part of the 
resolvent in Figs. 4 and 6 are consistent with this enhancement factor. 
\noindent\section{Conclusions}
We have studied the role of the fermion determinant in a random matrix model
at nonzero chemical potential. Previously, it was shown by Stephanov that 
the quenched version of this model is given by the limit $N_f \rightarrow 0$
of this partition function with the determinant replaced by its absolute 
value, and that, for example, the chiral condensate differs from the unquenched
result. 
We have demonstrated that the equality of the valence and sea quark masses 
is of 
fundamental importance in arriving at a consistent partition function.
A saddle point analysis was shown to be  in complete agreement with  an
explicit calculation of the Yang-Lee zeros of the partition function which were
found to be located on a curve in the complex plane.
If the valence quark mass and the sea quark mass are different, 
the condensate can diverge if the mass is inside the domain of the 
eigenvalues. We have shown this analytically for the $U(1)$-model of 
Gibbs and numerically for the random matrix model. The reason for this 
divergence is that, in the chirally broken phase, the partition function 
vanishes like $\exp(-n\mu^2)$ in the thermodynamic limit.

While  we do not expect that the random matrix model considered 
here leads to quantitative results, we believe that some of the generic 
features we have observed might be present in QCD as well. 

{\bf Acknowledgements}
\vglue 0.4cm
This work was partially supported by the US DOE grant
DE-FG-88ER40388. We thank Robert Shrock for useful discussions and educating us
on to general properties of Yang-Lee zeros. Misha Stephanov, 
Melih  \c{S}ener and G\'{a}bor Papp are thanked for useful discussions. 
James Osborn is thanked for
pointing out the existence of multiprecision packages. 
We acknowledge
D.H. Bailey and NASA Ames for making their  multiprecision package available.

Finally, we wish to point out that at the time of submission of this 
paper we received a preprint by J. Feinberg and A. Zee \cite{zee} with 
minor overlaps with the present work. In particular, parts of sections 2,
3.1 and 3.2 are also discussed in \cite{zee}.

\end{document}